\documentclass[prc,superscriptaddress,showpacs,byrevtex,twocolumn]{revtex4}

\usepackage{graphicx}% Include figure files
\usepackage{dcolumn}% Align table columns on decimal point
\usepackage{bm}% bold math

\begin{document}

%\preprint{MSUCL xxx}

\title{Electron Capture Strength for $^{60,62}$Ni and $^{58,60,62,64}$Ni($p,n$)$^{58,60,62,64}$Cu reactions at 134.3
MeV}

\author{N. Anantaraman}
\affiliation{National Superconducting Cyclotron Laboratory\\
Michigan State University, East Lansing Michigan 48824}
\author{Sam M. Austin}
\email{austin@nscl.msu.edu}
\homepage{www.nscl.msu.edu/~austin}
\affiliation{National Superconducting Cyclotron Laboratory and Joint Institute for Nuclear Astrophysics (JINA), Michigan State University, East Lansing, MI 48824}
\affiliation{Department of Physics and Astronomy, Michigan State
University, East Lansing Michigan 48824}
\author{B. A. Brown}
\affiliation{National Superconducting Cyclotron Laboratory and Joint Institute for Nuclear Astrophysics (JINA), Michigan State University, East Lansing, MI 48824}
\affiliation{Department of Physics and Astronomy, Michigan State
University, East Lansing Michigan 48824}
\author{G. M. Crawley}
\affiliation{National Superconducting Cyclotron Laboratory\\
Michigan State University, East Lansing Michigan 48824}
\affiliation{Department of Physics and Astronomy, Michigan State
University, East Lansing Michigan 48824}
\author{A. Galonsky}
\affiliation{National Superconducting Cyclotron Laboratory\\
Michigan State University, East Lansing Michigan 48824}
\affiliation{Department of Physics and Astronomy, Michigan State
University, East Lansing Michigan 48824}
\author{R. G. T. Zegers}
\affiliation{National Superconducting Cyclotron Laboratory and Joint Institute for Nuclear Astrophysics (JINA), Michigan State University, East Lansing, MI 48824}
\affiliation{Department of Physics and Astronomy, Michigan State
University, East Lansing Michigan 48824}
\author{B. D. Anderson}
\affiliation{Department of Physics, Kent State University, Kent
Ohio 44242}
\author{A. R. Baldwin}
\affiliation{Department of Physics, Kent State University, Kent
Ohio 44242}
\author{B. S. Flanders}
\affiliation{Department of Physics, Kent State University, Kent
Ohio 44242}
\author{R. Madey}
\affiliation{Department of Physics, Kent State University, Kent
Ohio 44242}
\author{J. W. Watson}
\affiliation{Department of Physics, Kent State University, Kent
Ohio 44242}
\author{C. C. Foster}
\affiliation{Indiana University Cyclotron Facility, Indiana
University, Bloomington, Indiana 47405}

\date{\today}

\begin{abstract}
{\bf Background:}  The strength of electron capture for medium
mass nuclei has a significant effect on the evolution of
supernovae. There is insufficient knowledge of these strengths and very little data for important radioactive nuclei.\\
{\bf Purpose:}  Determine whether it is feasible to obtain EC strength from studies of $T_o+1$ excitations in ($p,n$) reactions, and whether this might yield information for radioactive nuclei.\\
{\bf Methods:}  Cross sections for the $^{58,60,62,64}$Ni($p,n$)$^{58,60,62,64}$Cu reactions were measured over the angular range of 0.3${^\circ}$ to 11.6${^\circ}$ at 134.3 MeV using the IUCF neutron time-of-flight facility.\\
{\bf Results:} The $T_o+1$ excitations in $^{60,62}$Ni were identified by comparison with inelastic proton scattering spectra, their $B$(GT) were extracted, and the corresponding electron capture rates in supernovae   were calculated. Data from the TRIUMF ($n,p $) experiments at 198 MeV were reanalyzed;  the electron capture rates for the reanalyzed data are in moderately good agreement with the  higher resolution ($p,n $) results, but differ in detail.  The possibility of future measurements with radioactive nuclei was considered.\\
{\bf Conclusions:} It is  possible to determine electron capture
strength from ($p,n$) experiments.  This approach may make it possible to obtain electron capture strength for radioactive nuclei by studying ($p,n$) reactions in inverse kinematics.
\end{abstract}


\maketitle

\section{Introduction}
Interest in allowed Gamow-Teller strength in medium-mass nuclei
($A=20-70 $) is related to unresolved issues concerning weak
strength in nuclear physics and astrophysics.  The bulk of the
electron capture (EC) strength in nuclei is not energetically
accessible to direct measurement, but can be obtained from charge
exchange reactions (CER): the CER cross section at low momentum
transfer (small angles) is proportional to the Gamow Teller
strength $B$(GT) for sufficiently high bombarding energies, above
about 100-120 MeV/nucleon. The first systematic CER  electron capture strength studies were performed at TRIUMF \cite{alford98} using the ($n,p$) reaction at
$E_n \approx 200$ MeV and achieved a resolution of about 1 MeV  for a number of nuclei.

Large basis shell-model calculations for these
nuclei \cite{caurier99} are in reasonably good agreement with the
($n,p$) data. There are, however, significant differences for the
Ni isotopes \cite{caurier99,elkateb94,williams95}.  More recent work with higher resolution, mostly with the ($d$, $^2$He) reaction
\cite{frekers04}, has also been in general agreement with shell model calculations, but in some cases there are significant
differences even in centroid locations \cite{frekers05}. Data from
a recent $^{58}$Ni($t$,$^3$He)measurement \cite{cole06} agree with
the ($d$, $^2$He) results \cite{hagemann05} but less well with the
TRIUMF data. The ($t$,$^3$He) data are in good agreement with
shell model calculations using the KB3G two-body interaction at
low excitation energies $E_x$, but the agreement is poorer at high
$E_x$.  The converse is true for the GXPF1 interaction(for a
detailed discussion see Ref. \cite{cole06} and references therein).

These uncertainties in predicted EC strength introduce
uncertainties in predictions of the evolution of massive stars and
the ensuing core-collapse supernovae.  They also
affect nucleosynthesis in Type Ia supernovae and the crust
properties of neutron stars in accreting
binary systems. For details see Ref. \cite{langanke03} and references
therein. It appears that further experimental and theoretical work is necessary to better define the effective interactions used in
shell-model calculations and to permit more reliable calculations
of electron capture strength for astrophysical applications.

\begin{figure}
\includegraphics[width=7.5cm]{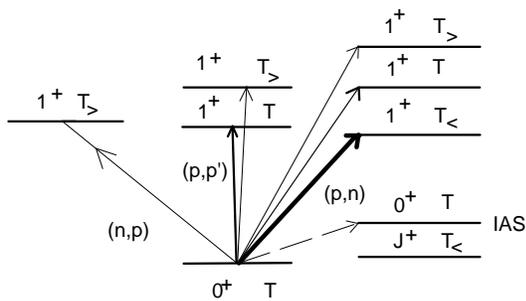}
\caption{Diagram of transitions via ($p,n$), ($n,p$) and ($p,p'$)
interactions.  More intense transitions are shown by darker lines.  With the exception of the transition to the isobaric analog state, those shown involve transfer of total angular momentum, spin, and isospin $\Delta J = \Delta S = \Delta T = 1$. States labelled with the same quantum numbers are isobaric analogs.   The symbols $T_>$, $T$, and $T_<$ stand for $T_o+1$, $T_o$ and $T_{o}-1$.  We are concerned here with the
relatively weak transitions to the $1^+, T_>$ states.}
\label{isfactor}
\end{figure}

In this paper we describe a less direct
approach to studying electron capture strength: obtaining $\beta
^+$ strength from studies of charge exchange in the $\beta ^-$
direction. The strength of $\beta ^-$ transitions to $T_o +1$
states in the residual nucleus, where $T_o$ is the isospin of the
target nucleus, is related by an isospin geometry factor to $\beta
^+$ strength from the same nucleus, as shown in Fig.\ \ref{isfactor}.
Specifically, $\beta ^+/\beta ^-=(T_o+1)(2T_o+1)$. This method
has been exploited previously for the obvious case of
self-conjugate nuclei and for the $T = 1$ nuclei $^{26}$Mg
\cite{madey87,zegers06} and $^{58}$Ni \cite{fujita96,fujita02,fuj07}. Here we apply the technique to nuclei with higher isospin: $^{60,62,64}$Ni.

To obtain $T_o +1$ strength with ($p,n$) reactions, one has to
deal with two important issues. First, the $T_o + 1$ states appear
at high excitation energy, and lie on a large background; this is
presumably the reason earlier experiments in this mass region with
 poorer statistics had not seen these states
\cite{rapaport83}. And second, charge exchange reactions such as
($p,n$) do not have an isospin meter; they are not selective of
isospin. In the present high statistics ($p,n$) experiment on the
$T=2$ and $T=3$ nuclei $^{60,62}$Ni, we observe  peaks at the
expected energies of $T_o +1$ states and present reasonably
convincing evidence that they are $T_o+1$ states. It, therefore,
appears that both of the above issues can be dealt with although
some ambiguities remain.

In Section II we  describe the experimental procedures and in Section III  present the results for $B$(GT) and reaction rates. In Section IV we discuss  some options for future measurements of electron capture strength for radioactive nuclei using inverse kinematics, with heavy radioactive beams incident on hydrogenous targets.

\section{Experimental Procedures}
\begin{figure}
\includegraphics[width=8cm]{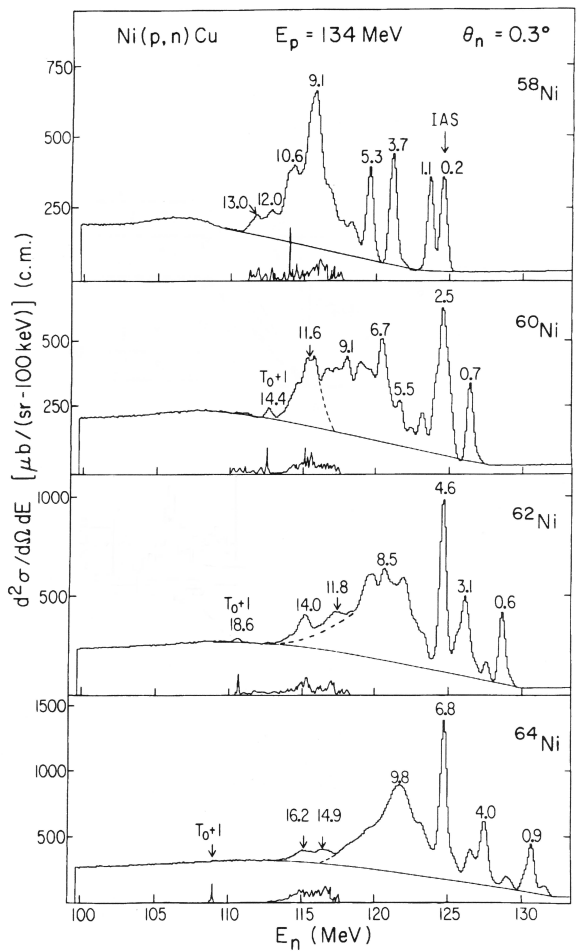}
\caption{Spectra for $^{58,60,62,64}$Ni ($p,n$) reactions at 134.4
MeV. There are about $10^4$ counts per channel in these spectra,
sufficient to observe the weak $T_o +1 $ states as described in
the text. The numbers above the peaks in the spectra are
excitation energies. Spectra observed in ($p,p'$) reactions \protect \cite{marty83} on the target nuclei are plotted on the energy axis.  The sharp peak at the left of each ($p,p'$) spectrum is the lowest-lying $T_o + 1$ state.} \label{nispectra}
\end{figure}

The beam-swinger time-of-flight system at the Indiana University
Cyclotron Facility was used to measure neutron time of flight
spectra resulting from the bombardment of 36 to 50 mg/cm$^2$
$^{58,60,62,64}$Ni targets (isotopically enriched to $>96\%$) by
134.3 MeV protons.  The detection station was placed at $0^{\circ}$,
85.8 m from the target, and consisted of three identical, large
volume, mean timed NE-102 detectors with a combined frontal area
of 1.55 m$^2$ and a thickness of 10.16 cm \cite{madey83}.  Data
were obtained at outgoing neutron angles of 0.3, 3.9, 8.0, and
11.6 deg for several different thresholds;  all thresholds gave
consistent results to within  $\pm 5\%$.  Efficiencies were
calculated with the Monte Carlo code of Cecil {\em et al.}
\cite{cecil79}.  The overall energy resolution was about 500 keV FWHM, worse than usual, because of an unusual and
uncorrectable jitter in the cyclotron timing signal.  The total
systematic uncertainty in the cross sections is $\pm 13\%$. The
peaks of main interest, see Fig.\ \ref{nispectra}, are those labelled
$T_o +1$, and located near $E_x = 14.4$ and 18.6 MeV in $^{60}$Cu
and $^{62}$Cu, respectively. The state in $^{60}$Cu was apparently
not seen in lower statistics work on that nucleus at 120 and 160
MeV \cite{rapaport83}. A more detailed view of the $T_o+1$ cross
sections is shown in panels (a) and (c) of Fig. \ref{6062ni}.
These results show that a high statistics experiment can observe
the $T_o + 1$ states, even with relatively poor resolution.

\section{Results}
\begin{figure}
\includegraphics[width=5cm]{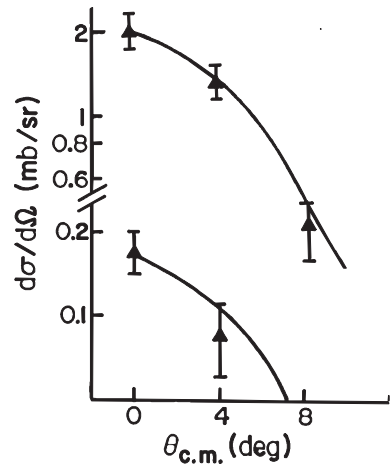}
\caption{Angular distributions for the $T_o+1$ excitations in
$^{60}$Ni (upper points) and $^{62}$Ni (lower points). The
$^{60}$Ni cross sections have been multiplied by 10 for display
purposes.  The curves are the DWBA calculations described in the
text.}
\label{angdist}
\end{figure}

The angular distributions for the high lying excitations in
$^{60,62}$Ni are shown in Fig.\ \ref{angdist}.  The curves are
DWBA calculated shapes for $L=0$ performed with the code DWBA70
\cite{raynal}, effective interactions at 140 MeV from Love and
Franey \cite{love81}, optical model potentials of Ref.
\cite{comparat74}, and simple ($\pi f_{7/2}\nu f_{5/2}^{-1}$) wave
functions.   The observed angular distributions are forward peaked
and are consistent with $L=0$ GT excitations.

We next deal with the evidence for assigning $T_o +1$ as the
isospin of these peaks.  Perhaps most important is the comparison
with spectra for the ($p,p'$) reaction that are shown on the energy
axis in Fig.\ \ref{nispectra}. The ($p,p'$) reaction near zero degrees
populates $1^+$ states preferentially, with a spin strength
proportional to $B$(GT) for the analog CER, but it can
populate both isospin $T_o$ and $T_o +1$ states.  The sharp states
seen at high excitation in the ($p,p'$) spectra have been assigned
as $T_o +1$ \cite{djalali82,marty83} for two main reasons.  First,
as $T_o$ of the target nucleus increases, these states shift
systematically to higher $E_x$ with respect to the $T_o$ strength,
as would be expected for a state of isospin $T_o +1$. And second,
although the states are unbound to neutron decay and have low
angular momentum, they are quite narrow; their observed width is
consistent with the experimental resolution, presumably
because the neutron decay of $T_o +1$ states is isospin forbidden
and the proton decay energy is well below the Coulomb barrier. The
positions of the sharp ($p,p'$) peaks agree approximately with those
seen in ($p,n$), after correcting for Coulomb effects, supporting
a $T_o+1$ assignment for the states seen in the ($p,n$) reaction.

Shell model calculations done in a simple basis \cite{brown} also
support the $T_o+1$ assignment; the $T_o +1$ strength is separated
from $T_o $ strength for $A > 58$, and is localized in a few
strongly populated states.  As expected, the separation grows as
$N-Z$ increases.  We shall see that these states lie low in the
spectra reached via ($n,p$) from the same target.

\subsection{Comparison with ($p,p'$)}
The isospin analog of a state at $E_x$(target), seen in the
($p,p'$) reaction, will occur in the ($p,n$) product nucleus at
approximately the same energy above the analog of the ground state (labeled IAS
in Fig.\ \ref{isfactor}), i.e., $E_x (p,n)=E_x (p,p')+E_x (IAS)$. In
Table \ref{energies} the relevant energies are tabulated, showing
that the energies of the analogs of the strongly excited ($p,p'$)
states and of the observed peaks in $^{60,62}$Cu agree within the
accuracy of the present measurements ($\pm 0.1$ MeV). There is
also a small enhancement at the expected energy in $^{64}$Cu (not
visible on the scale of Fig.\ \ref{nispectra}), but with the present
resolution, it is barely one standard deviation above
background, and is too weak to permit extraction of meaningful
cross sections. The observed widths of the lowest lying $T_o +1$ peaks are consistent with the resolution of the $(p,n)$ and ($n,p$)
experiments (these states are isospin forbidden to decay by
neutron emission and the proton decay energy for isospin allowed
decays is in the 2 to 3 MeV range, well below the Coulomb
barrier). The predicted excitation energies of the states in
$^{A}$Co that would be reached by the corresponding
$^{A}$Ni($n,p$) transitions are also given in Table
\ref{energies}.

\begin{table*}
\begin{minipage}{\textwidth}
\caption{Expected and observed energies of the lowest lying $T_o+1, 1^+$ states in
the Cu isotopes following ($p,n$) reactions and in the Co isotopes
following ($n,p$) reactions.
\label{energies}}
\begin{ruledtabular}
\begin{tabular}{ccccccc}
Target&$E_x(^{A}$Ni)\footnote{From the $^{A}$Ni ($p,p'$) results of Refs. \cite{djalali82,marty83}}&$E_x$(IAS)\footnote{Ref. \cite
{anthony88}}&$E_x (^{A}$Cu, expected)\footnote{From $E_x
(p,n)=E_x (p,p') + E_x (IAS)$}&$E_x (^{A}$Cu, observed)&$E_x
(^{A}$Co, predicted)\footnote{Calculated from
$E_x$($^A$Ni) in the first column and known Coulomb energies}&$E_x(^A$Co, observed)\footnote{From fits to the data of Ref. \cite{williams95} as shown in Fig. \ref{6062ni}.}\\
&(MeV)&(MeV)&(MeV)&(MeV)&(MeV)&(MeV)\\
\colrule
$^{60}$Ni & 11.85 $\pm$ 0.02 & 2.54 $\pm $0.02 & 14.39 & 14.4 $\pm$ 0.1 & 0.75&0.54\\
$^{62}$Ni & 14.00 $\pm$ 0.02 & 4.63 $\pm $0.01 & 18.63 & 18.6 $\pm$ 0.1 & 0.52&0.59\\
$^{64}$Ni & 15.62 $\pm$ 0.02   & 6.82 & 22.4 &                         & 0.25\\
\end{tabular}
\end{ruledtabular}
\end{minipage}
\end{table*}

\subsection{Determination of $B$(GT)}
We extracted the $B$(GT) corresponding to the $T_o +1$ excitations by
comparing their strength to that of the Fermi ($\Delta L=\Delta S=0)$ transition
to the IAS, ($B$(F) $\approx (N-Z)$), both evaluated at the same small
momentum transfer ($q \approx 0.05$ fm$^{-1}$) using the standard
techniques \cite{taddeucci87}. We make the usual assumption
\cite{taddeucci87}, fairly accurate for this energy range, that
the ratio of cross sections for Fermi and GT transitions of equal
strength is proportional to ($E_p(MeV)/54.9)^2$. This corresponds to a unit cross section, the ratio of cross section to $B$(GT), of 4.39 mb (4.23 mb) for $^{60}$Ni ($^{62}$Ni), in good agreement with the value of 4.49 mb (4.29 mb) used by \cite{williams95} at 198 MeV.  This is not surprising since the energy dependence of unit cross sections is weak.

\begin{figure}
\includegraphics[width=7.5cm]{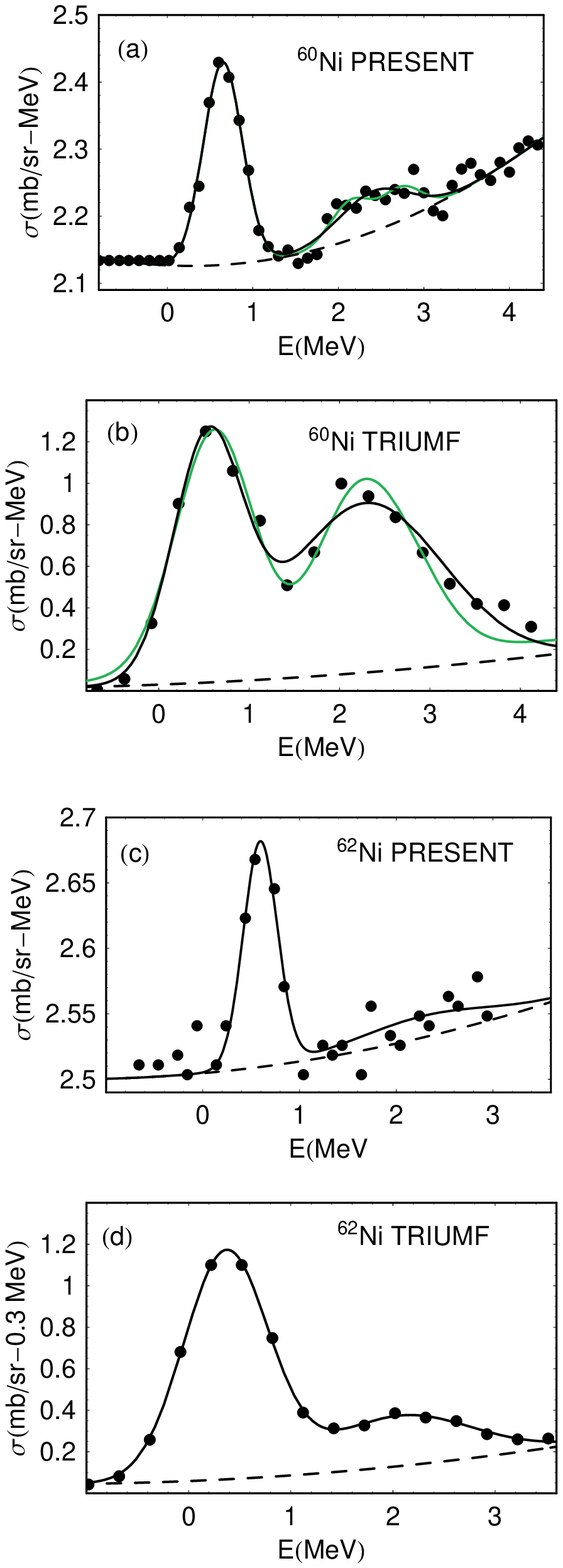}
\caption{(Color on line) Panel (a): Spectrum for $^{60}$Ni($p,n$)$^{60}$Co at 134.3 MeV, in the region of the $T_o+1$ states. The black and gray curves are two- and three-Gaussian fits, resp. Panel (b): Spectrum for $^{60}$Ni($n,p$)$^{60}$Co at 198 MeV. The black and gray curves are two- and three-Gaussian fits, resp.   Panel (c): Spectrum for
$^{62}$Ni($p,n$)$^{62}$Co at 134.3 MeV, in the region of the $T_o+1$
states. Panel (d): Spectrum for $^{62}$Ni($n,p$)$^{62}$Co at 198 MeV.  The 198 MeV data are from Ref. \cite{williams95} and the 134.3 MeV data are from the present work. Spectra are fitted with a second order polynomial background and two or three Gaussian peaks. For details see the text.} \label{6062ni}
\end{figure}

Determination of the number of counts in the $T_o + 1$ states in
$^{60,62}$Ni was done by fitting the data
with a polynomial background and a sum of Gaussians. The results
for a quadratic background and two or three Gaussians are shown in Fig.\ \ref{6062ni}. For $^{60}$Ni a linear background did not adequately
reproduce the overall spectrum shape. Fits to the $^{60}$Ni data with three Gaussian
peaks were not superior within statistics, although they allowed for the use of the same width, consistent with the energy resolution, for the three peaks. For $^{62}$Ni the lower
lying peak was well defined, but the strength of a second peak
could not be determined unambiguously; its area was fixed at the
same value relative to the lower excitation peak as in the TRIUMF
data (see below). This yields a satisfactory description of the
data as shown in Fig.\ \ref{6062ni}. The results for $B$(GT) are collected in columns (2) and (3) of Table \ref{bgt}. The uncertainties shown include a 13\% systematic error, dominated by the uncertainties in the cross section of the IAS (8\%) and in the extrapolation to $q=0$ (10\%).  In most cases the statistical uncertainty is larger, because the peaks sit on a large background.  The
$B$(GT) are converted to those that would be measured in $(n,p)$
reactions by multiplying by the appropriate ratio of
Clebsch-Gordon coefficients, 15 for $^{60}$Ni and 28 for
$^{62}$Ni.

\subsection{Comparisons with other ($p,n$) and ($n,p$)data}
Data for $^{58,60}$Ni are available from 120 MeV ($p,n$) measurements at
IUCF \cite{rapaport83}, but the statistics are not sufficient to
observe the weakly excited $T_o+1$ states in $^{60}$Ni. The 198
MeV Ni($n,p$) reaction studies at TRIUMF observe the
$^{58,60,62,64}$Ni EC states directly \cite{elkateb94,williams95}
and warrant a detailed comparison with the present data for
$^{60,62}$Ni. This is not possible for $^{58,64}$Ni.
The $T_o +1$ strength in $^{58}$Ni($p, n$) is not sufficiently separated
from $T_o$ strength to permit a reliable identification without
further information \cite{fujita96,fujita02,fuj07}. And as noted above
the strength we observe for $^{64}$Ni is significant only at the
one standard deviation level.

\begin{figure}
\includegraphics[width=7.5cm]{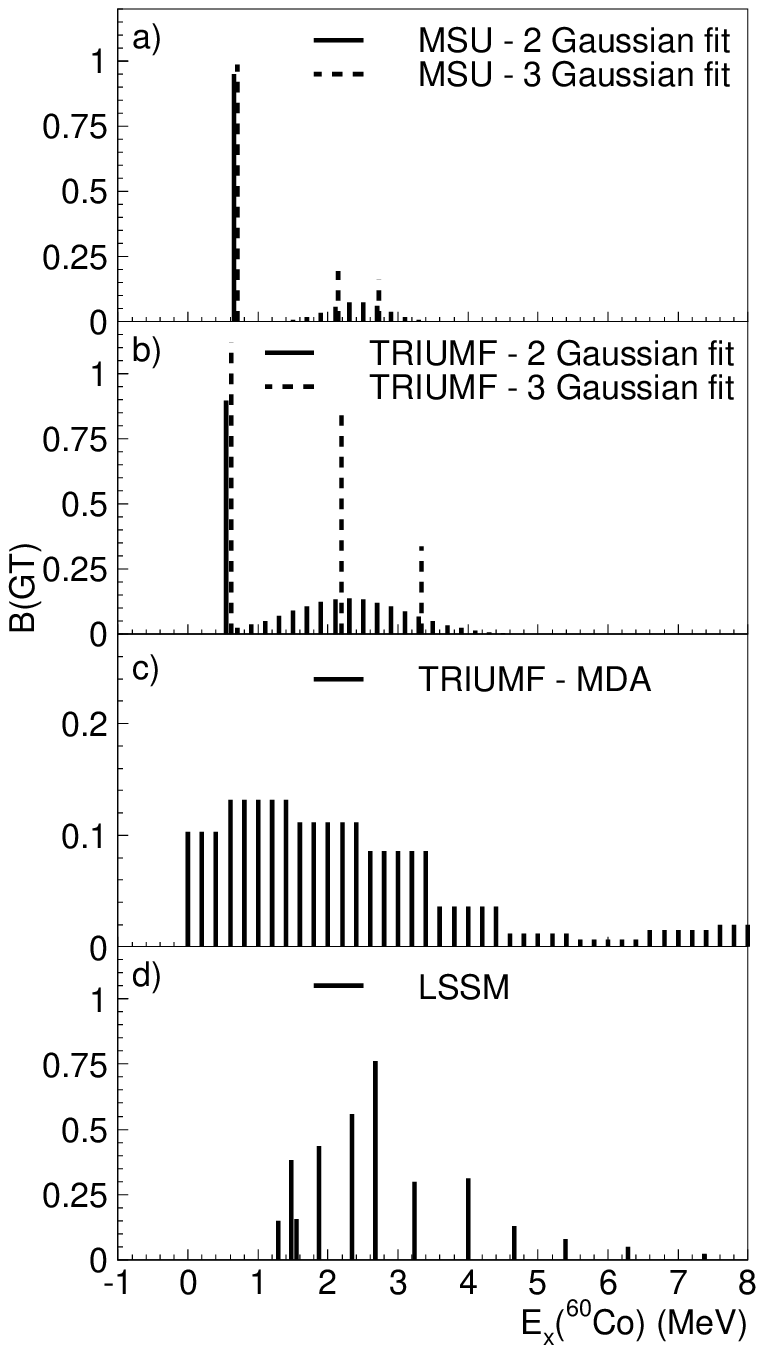}
\caption{Values of $B$(GT) for $^{60}$Ni obtained from the present 134.3 MeV ($p, n$) data and the 198 MeV TRIUMF ($n, p$) data. Panels (a) and (b) show the results of the fits to the ($p,n$) and the ($n, p$) cross sections as described in the text and shown in Fig, \ref{6062ni}.  Panel (c) shows $B$(GT) from the Multipole Decomposition Analysis (MDA) performed in Williams \cite{williams95}, and shown in Williams Fig. 12. The results from large scale shell model calculations \cite{caurier99} are shown in panel (d). For details see the text.}
\label{comparegt}
\end{figure}

Since only the stronger low-lying excitations can be extracted from the present data, in the 6th column of Table \ref{bgt} we compare with the results from ref.
\cite{williams95} as reported in Fig. 12 of that paper, integrated
over a comparable energy range, namely up to $E_x =$ 3.2 or 4.0 MeV.  The numbers quoted in Fig. 12 of Ref. \cite{williams95} in this
energy interval are about 25\% smaller than those given in Fig.
10 and Table II of that paper, as has been previously noted in
Ref. \cite{caurier99}; we use the results in Fig. 12 because they
are given as a function of excitation energy.

\begin{table*}
\begin{minipage}{\textwidth}
\caption{\label{bgt} Values of $B$(GT) for transitions to 1$^+$, $T_o +1$ states in $^A$Cu: ($p,n$); and in $^A$Co: ($n,p$). The values are those for two-Gaussian fits, except for the TRIUMF ($n, p$) results for $^{60}$Ni where the results for the three-Gaussian fits are shown in parentheses.}
\begin{ruledtabular}
\begin{tabular}{cccccc}
Target($E_x$-MeV)\footnote{The $E_x$ are the positions that these states would occur in the analog system $^{60}$Co.}&$B$(GT)-($p,n$)
&$B$(GT)-($n,p$)\footnote{Obtained by multiplying the results
obtained from ($p,n$) listed in the second column by the
isospin geometry factors:  15.0 for $^{60}$Ni, and 28.0 for
$^{62}$Ni.}&$B$(GT)-($n, p$)\footnote{From the data of Figs. 2 and 3 in
Ref. \cite{williams95} as analyzed in the present paper using, mainly, two-Gaussian fits. The values in parentheses for $^{60}$Ni are the results of three-Gaussian fits after summing the strengths for the upper two states; for separate values see Fig. \ref{comparegt}. For details see text.} &$B$(GT)$_{sm}$\footnote{From Caurier, {\sl et al.}
\cite{caurier99}. The strength quoted is the sum of
strengths to  $1^+$ states lying below 3.2 (4.0) MeV.}&$B$(GT)$_{TRIUMF}$\footnote{From Ref. \cite{williams95}, Fig. 12, integrated over the energy range up to 3.2 (4.0) MeV.} \\
\colrule
$^{60}$Ni(0.65)&0.063 $\pm$ 0.010 &0.95 $\pm$ 0.15&0.89 (1.03)&&\\
$^{60}$Ni(2.4)&0.026 $\pm$ 0.008&0.39 $\pm$ 0.12&1.47 (0.97)&&\\
$^{60}$Ni(0.65+2.4)&0.089 $\pm$ 0.014&1.34 $\pm$ 0.22&2.36 (2.00)&2.7 (3.0)&2.0\\
$^{62}$Ni(0.6)&0.032 $\pm$ 0.007&0.89 $\pm$ 0.20&1.01&&\\
$^{62}$Ni(2.3)&0.014 $\pm$ 0.005&0.39 $\pm$ 0.14&0.40&&\\
$^{62}$Ni(0.6+2.3)&0.046 $\pm$ 0.010&1.28 $\pm$ 0.29&1.41&1.9 (2.0)&1.3\\
\end{tabular}
\end{ruledtabular}
\end{minipage}
\end{table*}

A somewhat more detailed qualitative comparison is also possible.
Figs. 2-4 of Ref. \cite{williams95}, referred to as "Williams" in the following discussion, present the ($n,p$) data prior to multipole decomposition.  Those data are given in smaller bins than  the final results, 300 keV compared to 1.0 MeV, and have structure that did not survive the multipole decomposition procedure. For example, the spectrum for $^{60}$Ni in Fig. 2 of Williams has two peaks below $E_x = 4$ MeV that, as is shown by their Fig. 6, are dominated by $L=0$ strength.  These peaks are not separately visible in the $L=0$ spectra of Williams, Fig. 9.
We have scanned the Williams data (in their Figs. 2 and 3) for $^{60,62}$Ni and fitted them with quadratic backgrounds and two or three Gaussians as was done for the 134.3 MeV data. For the Williams $^{60}$Ni data, the three-Gaussian fits were superior.  The quadratic backgrounds presumably account mainly for the contribution of higher $L$ transitions, at least for the Williams data.

 In Fig.\ \ref{6062ni} we show both the Williams cross section data at $E_n=198$ MeV and the present data for the $T_o+1$ states at $E_p=134.3$ MeV. For $^{60}$Ni, the locations and spacing of the two lowest states are in excellent agreement; for $^{62}$Ni the position of the lowest lying state is the same  within about 170 keV, consistent with combined experimental uncertainties. The cross sections for the lowest lying states near 0.6 MeV agree within the uncertainties.  However, the relative intensities of the two lowest states for $^{60}$Ni observed in the  present 134.3 MeV data differ significantly from those in the Williams data. The reason for this difference is not understood.  We have investigated whether changes in the details of the fitting procedure could significantly change this ratio;  systematic changes in the ratio of more than 15\% seem unlikely.

 In order to convert the Williams cross section data of Fig. \ref{6062ni} to $B$(GT), the cross sections were extrapolated to $0^{\circ}$ using the $^{64}$Ni($n,p$) angular distribution shown in  Williams, Fig. 5, and then to $q=0$ using the momentum transfer dependence found in the present 134.3 MeV data. The unit cross sections from Williams were used to convert the resulting cross sections to $B$(GT).   The results are shown in Column 4 of Table  \ref{bgt}.  Only statistical errors, typically 3-5\% are quoted in Williams.  It seems probable, however, given various experimental uncertainties and uncertainties in the unit cross section that the overall uncertainties are at least 10\% and perhaps larger.

 As we have noted for the cross sections, the values of $B$(GT) from the ($p,n$) and ($n,p$) reactions  agree within the uncertainties  for the states near 0.6 MeV  in $^{60,62}$Ni, but the excitation of the 2.4 MeV state in $^{60}$Ni is much stronger in ($n,p$).

\subsection{Comparisons with shell-model calculations}
In Table \ref{bgt} we compare our results with the large basis
shell-model calculations of Ref. \cite{caurier99}. These
calculations use a renormalized (reduced) GT operator, $g_A/g_V
=1.0$  The theoretical strength is concentrated at low excitation,
mostly below $E_x= 4.0$ MeV.  However, we shall see later that the energy distribution of theoretical strength for $^{60}$Ni differs greatly from experiment, mainly lying near the high energy peak of Fig. \ref{6062ni}.

\subsection{Comparisons of $B$(GT) and electron-capture rates}
The values of $B$(GT) for $^{60}$Ni from the two experiments and from the shell model calculations of Ref. \cite{caurier99} are shown in Fig. \ref{comparegt}.  In cases where the fitted cross section peaks had widths consistent with the experimental resolution we plotted their $B$(GT) at the position of the peak;  this applied to the lower lying peak in all the data fitted in this paper and to the higher-lying peak when fitted by the sum of two Gaussians.  When the higher-lying peak was fitted by a single Gaussian, the fitted width was greater than the resolution, and the strength was divided into 200 keV bins. The values obtained are plotted in panels (a) and (b) of Fig. \ref{comparegt}. In Fig. 12 of Williams \cite{williams95} the $B$(GT) are given in 1.0 MeV wide bins, with one of the bins extending to $E_x$ = -0.5 MeV. The 1.0 MeV bins were subdivided into 200 keV intervals, and the values at $E_x < 0$ MeV were incorporated uniformly into the lowest three bins;  the results are shown in panel (c). The shell model results of Ref. \cite{caurier99} are shown in panel (d).

\begin{center}
\begin{figure*}
\includegraphics[width=15cm]{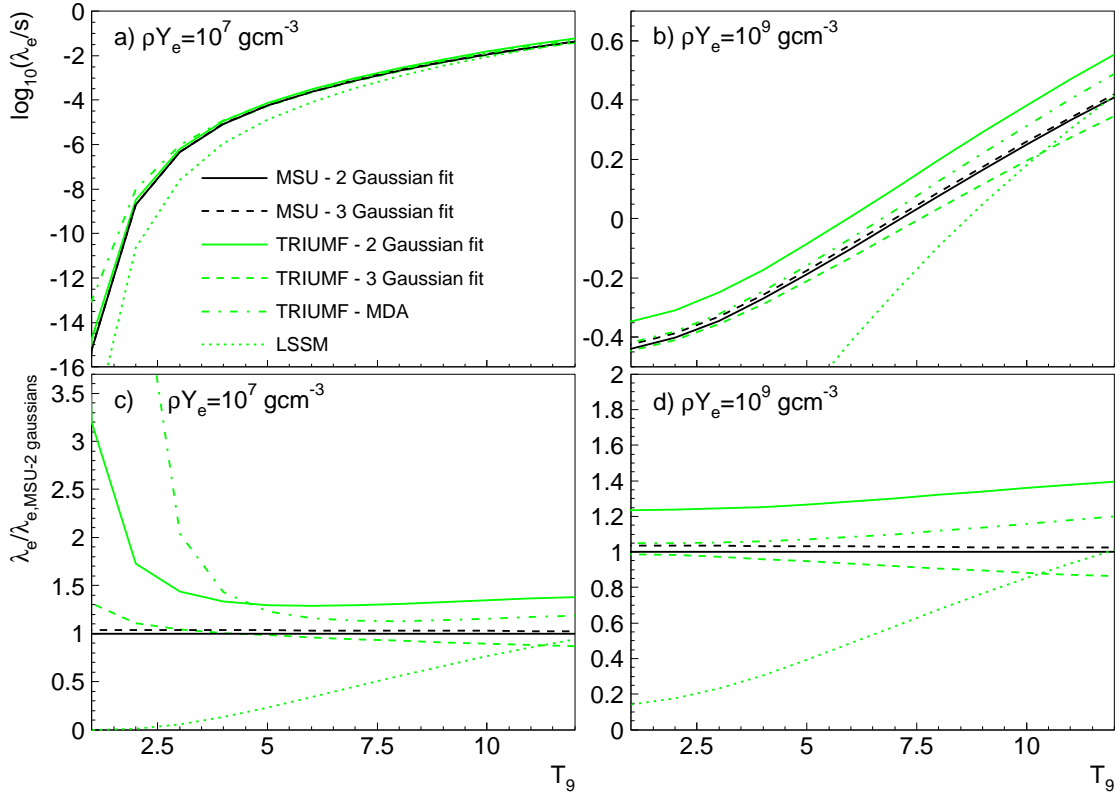}
\caption{(Color on line)Reaction rates obtained from the $B$(GT) of Fig. \ref{comparegt}. The upper two panes show the absolute rates and the lower two panels the results compared to those for the 134.3 MeV ($p,n$) data. For details see the text.}
\label{capturerates}
\end{figure*}
\end{center}

There are significant differences among the distributions of $B$(GT) of Fig. \ref{comparegt}.  The results from Fig. 12 of Williams \cite{williams95} extend to lower $E_x$  than do those from the analyses of the same date carried out in this paper, presumably as a result of the binning procedure used in the multipole analysis.  This will result in larger electron capture rates at relatively low temperatures and densities in astrophysical environments, as we show in Fig. \ref{capturerates}. We would argue that the results of the present analysis are more reliable. On the other hand, the strengths predicted by the shell model \cite{caurier99} are peaked at high energies and will become important only at rather high temperatures or densities.

Electron capture rates are calculated for the different distributions of Fig. \ref{comparegt} using the code described in Ref. \cite{bec06}. Electron chemical potentials were computed from a tabulation \cite{tim00}. These calculations ignore contributions from higher-lying states and from capture on thermally excited states that will be important at high temperatures and densities. Details of the calculation and additional references  are given in \cite{cole06}.

Rates were calculated on a grid including values of $\rho Y_e$ from $10^1$ to $10^{14}$ gcm$^{-3}$ and of $T$ from (0.01-100) $\times 10^9$ K. In Fig. \ref{capturerates} we show the rates for two representative $\rho Y_e$  of interest in the pre-supernova evolution of massive stars:  $\rho Y_e = 10^7$ and $10^9$ gcm$^{-3}$.  In a 25$M_{sun}$ star, for example, the former is characteristic of various stages of Si burning, when the temperature $T_9 \approx 2-4$; the latter occurs late in the pre-supernova stage when $T_9 \gtrsim 10$. In panels (a) and (b) the absolute rates are shown: in panels (c) and (d) the rates are compared to those for the present 134.3 MeV data. For details, see the text.

We find that: (1)  The rates for the two- and three-Gaussian fits to the 134.3 MeV ($p, n$) data are almost the same over the entire parameter space, reflecting the similarity of the two and three-Gaussian fits.
(2) The TRIUMF results for the Gaussian fits made in the present paper differ from the 134.3 MeV data, but rates for the three-Gaussian fits are in fairly good agreement, especially at the higher densities and/or temperatures. (3) At lower temperatures and densities, the rates for the $B$(GT) results presented in Fig. 12 of Williams \cite{williams95} are considerably  higher than the others shown.

We conclude from these comments that the Multipole Decomposition Analysis for the TRIUMF data affects the resulting $B$(GT)s significantly, at least for $^{60}$Ni.  It also appears that obtaining electron capture rates from ($p, n$) data is a viable procedure.

\section{Measurements of EC strength for radioactive nuclei.}

Studies of the electron capture strength of radioactive nuclei must be done using inverse kinematics (IK), with a radioactive beam of heavy nuclei incident on a light target. Such studies will be necessary to explore nuclei with significant neutron excesses and to make possible studies on odd-odd nuclei;  among these only $^{50}$V is a stable target.
Under the  conditions of interest ($\theta_{c.m.}$ near $0^{\circ}$, $E_x$ in the 0-15 MeV range) the outgoing light particles typically have small energies;  if these particles are charged, exceedingly thin targets are required, yielding a very small reaction rate. The ($p,n$) reaction does not have this problem; the low energy neutrons can easily leave the target.

There are, however, limitations on the use of the ($p,n$) technique. It is applicable only to nuclei with isospin large enough  that the splitting of $T_o $ and $T_o +1$ states allows one to isolate $T_o +1$ strength with reasonable certainty. For a $T_o =1$ nucleus like $^{58}$Ni the $T_o $ and $T_o +1$ excitations are strongly intermixed as discussed in detail in ref. \cite{fujita96}. Because the strength of a transition  is roughly
proportional to $1/T^2_o$, the isospin  must also be sufficiently
small that the $T_o +1$ states are observable.  In the present
experiment the $T_o+1$ states were barely seen in $^{64}$Ni with
$T_o=4$.  Better resolution would increase the peak to background
ratio and make it possible to observe $T_o +1$ states for nuclei
with higher isospin. If, for example,  a resolution of 200 keV could be obtained for IK ($p,n$) reactions, 2.5 times better than in the present experiment, one could study nuclei where the relative strength of the
$T_o + 1$ excitations is a factor of 2.5 smaller, corresponding to
$T_o$ as large as 5 ($^{62}$Ni has $T_o = 3$). It is not clear whether such nuclei can be reached in practice; one may be limited by the intrinsic decay widths or spreading widths of the states.  And obtaining 200 keV resolution will be challenging; it will certainly require the intensities of an advanced radioactive beam facility.  At  present  intensities feasible resolutions are in the 0.5-1.0 MeV range.

Inverse kinematics ($p, n$) approaches are being undertaken at the NSCL.  Detecting the low energy neutrons is feasible, but presents a significant challenge. Moreover, the c.m. energy typically depends on the laboratory angle of the emitted neutron so the detection system must have high angular granularity to obtain good resolution in $E_x$.  Construction of a detector that will meet these challenges is underway.

\section{Summary}
We have shown that ($p,n$) reactions at 134.3 MeV have sufficient sensitivity to extract $B$(GT) for strongly excited $T_o + 1$ states,
provided that the isospin of the target nucleus is neither too
large nor too small. Electron capture strengths for the lowest
lying $T_o +1$ states in $^{60,62}$Ni were extracted from data for
the ($p,n$) reaction and compared with ($n,p$) data and with large
basis shell model predictions.  The fits to the raw TRIUMF ($n,p$) cross section data performed in the present paper yield results rather close  to the ($p, n$) results for the lowest lying peak but have larger strength to higher-lying states for $^{60}$Ni as shown in Fig. \ref{6062ni}.  However, the multipole analysis leading to the $B$(GT)  presented in Williams Ref. \cite{williams95} moves some strength to lower energies.

Electron capture rates were calculated for two cases of interest
in the pre-supernova evolution of massive stars.  The results for the present analyses of the 134.3 MeV ($p,n$) data and of the Williams ($n, p$) data are in reasonable agreement, except at the lowest temperatures and densities.  However, the $B$(GT) extracted by the Williams multipole analysis yields significantly larger rates, particularly at low $T$.

That ($p, n$) reactions lead to electron capture rates that are in agreement with ($n, p$) results, when both are analyzed in the same fashion,  supports using ($p,n$) reactions in IK to study $T_o+1$ states in radioactive nuclei.  We conclude that the IK approach may be useful for the radioactive nuclei that play an important role in supernova evolution and whose electron capture strength is difficult to obtain in other ways.

\begin{acknowledgments}
The authors wish to thank Parker Alford, Y. Fujita, M. Fujiwara,
and Grant Mathews for useful conversations and comments. This work
was supported in part by the National Science Foundation.
\end{acknowledgments}

%\bibliography{bet3sub1}

\end{document}